\RequirePackage{amsmath}
\documentclass[runningheads]{llncs}
\usepackage[T1]{fontenc}

\usepackage{graphicx}

\usepackage{hyperref}
\usepackage{multicol}
\usepackage{multirow}
\usepackage[most]{tcolorbox}
\usepackage{xcolor, soul}
\usepackage{ctable}
\usepackage{subfig}
\usepackage{adjustbox}
\usepackage{array}

\usepackage{acronym}
\usepackage{enumitem}   
\usepackage{csquotes}
\usepackage{mdframed}
\usepackage{float}

\acrodef{SMS}[SMS]{Systematic Mapping Study}
\acrodef{RQ}[RQ]{Research Question}
\acrodef{ML}[ML]{Machine Learning}
\acrodef{ML-based}[ML-based]{Machine Learning-based}
\acrodef{MLOps}[MLOps]{Machine Learning Operations}
\acrodef{WF}[WF]{Workflow}
\acrodef{RA}[RA]{Reference Architecture}
\begin{document}

\title{An Analysis of MLOps Architectures: \\ A Systematic Mapping Study}

\titlerunning{An Analysis of MLOps Architectures}

\author{Faezeh Amou Najafabadi\inst{1}\orcidID{0009-0006-6413-9118}  \and
Justus Bogner\inst{1}\orcidID{0000-0001-5788-0991} \and
Ilias Gerostathopoulos\inst{1}\orcidID{0000-0001-9333-7101} \and
Patricia Lago\inst{1}\orcidID{0000-0002-2234-0845}}
\authorrunning{F. Amou Najafabadi et al.}

\institute{Vrije Universiteit Amsterdam, The Netherlands 
\email{\{f.amou.najafabadi,j.bogner,i.g.gerostathopoulos,p.lago\}@vu.nl}}

\maketitle     

\begin{abstract}
\textit{Context}. 
Despite the increasing adoption of Machine Learning Operations (MLOps), teams still encounter challenges in effectively applying this paradigm to their specific projects. 
While there is a large variety of available tools usable for MLOps, there is simultaneously a lack of consolidated architecture knowledge that can inform the architecture design.
\noindent \textit{Objective}. 
Our primary objective is to provide a comprehensive overview of (i) how MLOps architectures are defined across the literature and (ii) which tools are mentioned to support the implementation of each architecture component.
\noindent \textit{Method}. 
We apply the Systematic Mapping Study method and select 43 primary studies via automatic, manual, and snowballing-based search and selection procedures.
Subsequently, we use card sorting to synthesize the results.
\noindent \textit{Results}. 
We contribute (i) a categorization of 35 MLOps architecture components, (ii) a description of several MLOps architecture variants, and (iii) a systematic map between the identified components and the existing MLOps tools. 
\noindent \textit{Conclusion}.
This study provides an overview of the state of the art in MLOps from an architectural perspective. 
Researchers and practitioners can use our findings to inform the architecture design of their MLOps systems. 

\end{abstract}
\keywords{Machine Learning Operations \and MLOps \and Architecture \and Components \and Tools \and Systematic Mapping Study.}

\section{Introduction}
The use of \ac{ML} continues to grow in industry, and developing high-quality \ac{ML} models is important to sustain it.
However, when creating ML-based systems, another major concern of \ac{ML} engineers and operations teams is the effective deployment and maintenance of \ac{ML} models in production~\cite{kreuzberger_machine_2023}. 
To address this, the \ac{MLOps} paradigm has formed in industry~\cite{John2021}.
Similar to DevOps~\cite{Bass2015}, \ac{MLOps} comprises a set of best practices and related methods, technologies, and tools that aim to bridge the gap between the development of ML models and their deployment, maintenance, and evolution.
Despite the increasing adoption of \ac{MLOps}~\cite{ref_introducingMLOps_oreilly_2021}, it is still challenging for practitioners to effectively apply the paradigm to their projects~\cite{kolar2023barriers_MLOps}. 
First, there is a large variety of available tools usable for MLOps, which makes it hard for practitioners to analyze and compare all the options at their disposal~\cite{idowu_asset_2021}.
Second, while reusable design decisions for certain parts of ML-enabled systems are starting to emerge~\cite{warnett_architectural_2022}, there is still a lack of consolidated MLOps architecture knowledge that could guide architectural decisions. 
Third, MLOps evolves at a fast pace and simultaneously in different domains, which makes it difficult to discern the generalizable concepts and technologies from the domain-specific ones. 

In this paper, we therefore aim to provide an overview of the state of the art in \ac{MLOps} from an architectural perspective. 
We extract and analyze the components that comprise typical architectures of MLOps systems and identify several variants of such architectures based on the existing variability points, e.g., the optional presence of an online training pipeline. 
We analyze the dependencies between components and synthesize them in the form of a UML component diagram. 
To align terminology in a rather scattered domain, we provide several known aliases for each component we identify. 
We also extract and analyze the tools that are mentioned in the MLOps literature and map them to the architecture components. 
This provides insight into the implementation options for each component and sheds light on untapped R\&D opportunities in the form of less tool-supported components. 

We accomplish this by performing a \ac{SMS} of the scientific literature of \ac{MLOps}, which grounds our analysis in the scientific state of the art. 
As such, we complement other attempts that analyzed gray literature for MLOps best practices and architecture design decisions~\cite{idowu_asset_2021,warnett_architectural_2022}. 
As our long-term goal, we aim to provide a comprehensive reference architecture~\cite{nakagawa2023reference-architecture} for MLOps that covers both the \textit{structural} perspective (the focus of this paper), and the \textit{process} and \textit{stakeholder} perspectives. 

In summary, the contributions of this study are (i) a synthesis and categorization of 35 MLOps architecture components, (ii) a description of several MLOps architecture variants, and (iii) a systematic map between the identified components and the existing MLOps tools. 

The target audience of our study is (i) researchers in software architecture for ML, who can build on our results to derive and consolidate architecture knowledge in the form of MLOps best practices and patterns, and (ii) ML practitioners, who can use our findings to inform the architecture design of their MLOps systems.  

\section{Related work}\label{sec_related_work}
\ac{MLOps} definitions, practices, and guidelines have been the subject of numerous secondary studies including scoping reviews, systematic literature reviews, and multivocal literature reviews. 
Many secondary studies on this topic aim at \textbf{clarifying the definition of MLOps}. 
Mboweni et al.~\cite{mboweni2022SLR_MLDevOps} state that there is still no official standard definition for MLOps.
Based on their systematic review to disambiguate the definition of MLOps in the literature, they claim that they did not find evidence of a common understanding among scholars and experts on how MLOps should be implemented and institutionalized across the industry to create a common vision.
Lima et al.~\cite{lima2022mlops_SLR} systematically reviewed 30 papers aiming at deriving practices, standards, roles, maturity models, challenges, and tools for MLOps. Based on the addressed challenges and assessment of models, they draw the conclusion that \enquote{research on MLOps is still in its initial stages.}

Some literature reviews provide methodologies for \textbf{effectively approaching MLOps projects}.
Testi et al.~\cite{testi2022SLR_mlops_taxonomy} provide a taxonomy of the current approaches toward and propose a methodology for addressing MLOps projects.
Kolltveit and Li~\cite{kolltveit2022operationalizingMLModels} specifically focus on the operationalization of ML models with regard to tools and infrastructure that are deployed in different stages of MLOps workflows.
Recupito et al.~\cite{recupito2022MLOpsTools_MLR} take a different perspective and provide an overview of the most common tools and their characteristics that support the creation of MLOps pipelines, without a clear mapping to components.

Several papers on the \textbf{architecture of ML-based systems} are also closely related to our study.
For example, Warnett and Zdun conducted two studies in which they used practitioner gray literature to synthesize architectural design decisions (ADDs) for ML workflows~\cite{Warnett2022} and ML deployment~\cite{warnett_architectural_2022}, with several of their sources being blog posts about MLOps.
As a result, several of their ADDs are related to architecture components that we synthesize in our study.
However, they do not combine this knowledge into a holistic architecture and also do not cover several parts of MLOps, such as inference and monitoring.
In a controlled experiment, Warnett and Zdun~\cite{Warnett2024} also compared the understandability of informal textual and graphical MLOps architecture representations with semiformal MLOps architecture diagrams.
They conclude that the understandability of MLOps architecture descriptions is significantly larger with supplementary semiformal architecture diagrams.
Lastly, Kumara et al.~\cite{kumara2023requirements_ra_mlops} strive towards a reference architecture of MLOps by eliciting requirements and components from the gray literature.
In their preprint, they provide a layered architecture that focuses on requirements that the MLOps environment needs to provide. 

Our own study complements the above studies by using scientific literature to synthesize MLOps architecture components, their relationships, and supporting tools to implement them.
The results of this synthesis address gaps identified by previous studies, namely \enquote{no common understanding of MLOps definition} and \enquote{no clear mapping between the tools and the related components}.
Moreover, unlike existing works, we also synthesize and discuss several architectural variants.
Researchers and practitioners can use the proposed architectures to identify a suitable variant for their requirements. 

\section{Methodology}\label{sec_methodology}
In this section, we describe the research goals and process of our study. We design and follow a rigorous protocol, by following established guidelines for systematic secondary studies~\cite{keele2007_SLRguidelines,petersen_2015_SMSguidelines}.

\subsection{Goal and research questions}\label{subsec_goal_rqs}

The main goal of this \ac{SMS} is to identify, classify, and analyze the architectures of existing MLOps systems described in scientific literature. In particular, we focus on their structural view, namely the individual components, their dependencies, and their responsibilities.
The target audience of this study is (i) practitioners, e.g., ML engineers and software architects, who need to obtain an overview of the common MLOps landscape to make informed decisions, and (ii) researchers aiming to improve the state of the art in architectures of MLOps systems. 
To approach this goal, we have phrased our overall \ac{RQ} as \textbf{\enquote{How are MLOps architectures described in the scientific literature?}}, and broke it down into the following sub-\ac{RQ}s:

\begin{enumerate}[leftmargin=1cm]
    \item[RQ1:] Which are the different components and their dependencies? \\
    \textit{This RQ helps us categorize the various components within an MLOps architecture and comprehend how different researchers and practitioners interconnect these components.}
    \item[RQ2:] Which tools are used to support or implement the identified components? \\
    \textit{This RQ helps us identify tools that can be used for implementing an architecture component, as well as identify the most and the least tool-supported components.}
\end{enumerate} 
       
\subsection{Research process} \label{subsec_study_design}

\begin{figure}[t]
    \centering
    \includegraphics[width=1.0\textwidth]{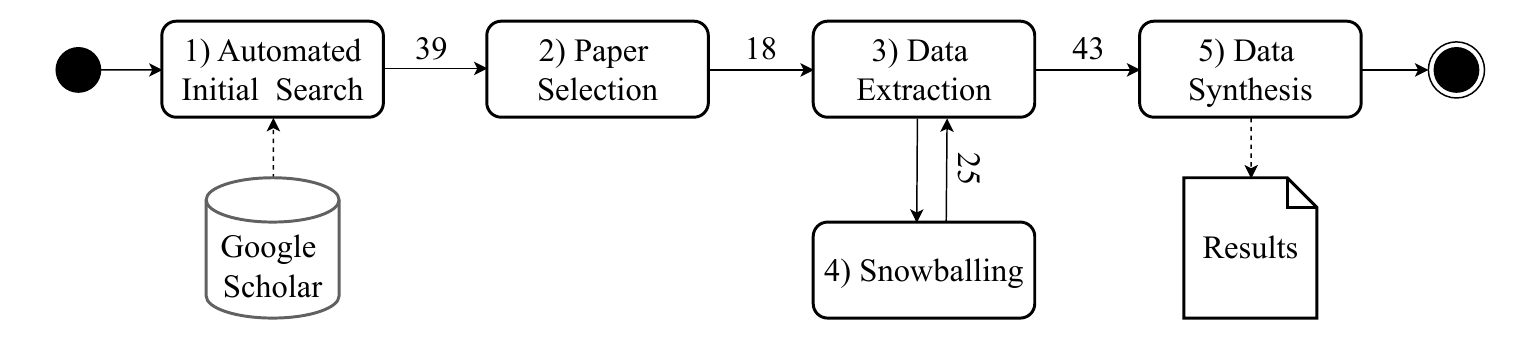}
    \caption{Overview of the research process}
    \label{fig_research_strategy}
\end{figure}

We visualize our research process in~Fig.~\ref{fig_research_strategy} and describe the different steps in this section. 
Essentially, we first obtain an initial set of papers via an automated title-based search on Google Scholar, then filter this to obtain a starting set of primary studies. After extracting data from the first set of primary studies, we augment this set via bi-directional one-step snowballing~\cite{Wohlin2014_snowballing} and extract data from the new set of primary studies.  
Using Google Scholar as a meta-engine allows us to avoid bias towards specific publishers~\cite{ampatzoglou2019threatsinSE}. 
Finally, the extracted data is used in a rigorous synthesis process.

\subsubsection{Step 1: Automated initial search}

By extracting relevant search terms from MLOps literature known to us and adding synonyms, we iteratively construct and test several search strings. As a result, we finally arrive at the following query to search in titles of the available literature:

\smallskip
\noindent\textit{allintitle: (MLOps OR \enquote{machine learning operations}) AND (model OR models OR pipeline OR pipelines OR architecture OR architectures OR architecting OR workflow OR workflows OR process OR processes)} 
\smallskip

The advantages of this final title-focused variant are its manageable number of results and the low number of false positives.
Additionally, its potential limitation regarding the extensiveness of its results is compensated by snowballing.
The automated query was executed via Google Scholar in September 2023 and yielded 39 potentially relevant papers. 

\subsubsection{Step 2: Paper selection}

The following inclusion (I) and exclusion (E) criteria are used during the selection of primary studies.
\vspace{-0.4pt}
    \begin{itemize}
        \item[I1] The paper describes the architecture of an MLOps system or part thereof.
        \item[I2] The described MLOps architecture is an original contribution of the paper, not simply cited related work.
        \item[E1] The paper is not written in English.
        \item[E2] The paper is a shorter or earlier version of a paper that is already included.
         \item[E3] The paper is a secondary or tertiary study.
         \item[E4] The full text of the paper is not available.
         \item[E5] The paper is not peer-reviewed and published.
    \end{itemize}

To be included in the list of primary studies, a paper has to fulfill all inclusion and no exclusion criteria.
To arrive at the above selection criteria, we conduct a selection pilot in which two reviewers independently review five papers and discuss their selection strategy in a consensus meeting. After fine-tuning the selection process, it is applied to all potentially relevant papers. Each paper is independently evaluated by two researchers for inclusion, with a consensus being necessary for final inclusion. 
Applying the selection criteria results in 18 papers that form our starting set used in the first round of data extraction and in snowballing. 

\subsubsection{Step 3: Data extraction} \label{subsec_extraction_strategy}

In this step, we systematically analyze the primary studies and extract data related to the \ac{RQ}s. 
To refine our data extraction strategy, we first conduct a data extraction pilot on five randomly chosen papers.  
Data from each paper is extracted by two authors independently and discussed in a consensus meeting leading to the final data extraction framework depicted in~Table~\ref{tbl_extraction}.
For RQ1, we extract the items \enquote{Architecture or process figures} and \enquote{Architecture components}, while we extract \enquote{Tools} and \enquote{Tool-component mapping} for RQ2.
The \enquote{Application domain} and the \enquote{Author affiliations} are extracted as generic information to use for further analysis.
From this point on, each of the remaining papers is assigned to two authors for data extraction. 
The extracted data is validated through bilateral discussions and consensus between the authors. 

\begin{table}[t]
\caption{Data Extraction Framework} \label{tbl_extraction}
\centering
\scriptsize
\begin{tabular}{p{0.08\textwidth}|p{0.3\textwidth}|p{0.58\textwidth}}
\specialrule{.1em}{.05em}{.05em}
\textbf{RQ} & \textbf{Data Item} &  \textbf{Notes}\\
\specialrule{.1em}{.05em}{.05em}
Generic & Application domain
& The respective domain the architecture is proposed for\\
\specialrule{.1em}{.05em}{.05em}
Generic & Author affiliations
& Origin of the architecture: academia, industry, or collaboration\\
\specialrule{.1em}{.05em}{.05em}
RQ1 & Architecture or process figures & The list of relevant figures and their types (values can be architecture, process, or combined)\\
\specialrule{.1em}{.05em}{.05em}
RQ1 & Architecture components
& The list of components and their relationship\\
\specialrule{.1em}{.05em}{.05em}
RQ2 &Tools
& The list of tools that are used or suggested in the paper\\
\specialrule{.1em}{.05em}{.05em}
RQ2 & Tool-component mapping
& The list of tools mapped to the components\\
\specialrule{.1em}{.05em}{.05em}

\end{tabular}
\end{table}

\subsubsection{Step 4: Snowballing}
We apply backward and forward snowballing to enrich the results obtained via automated search, as suggested by Wohlin et al.~\cite{Wohlin2014_snowballing}. 
During this step, all papers that either cite or are cited by a paper from the starting set are examined for inclusion in the final set of primary studies. 
We apply both the same selection criteria and data extraction (steps 2 and 3) as for the initial set of papers and the same process: each paper is examined independently by two researchers and consensus needs to be reached for including it, then the data from each paper extracted by two researchers and the results are discussed.  
After conducting a first round of bidirectional snowballing, we add 25 more papers to our set of primary studies. 
We limit the snowballing to a single round; this decision stemmed from the observation that subsequent data extraction yielded minimal additional components and tools compared to the initial seed collection.

\subsubsection{Step 5: Data synthesis} \label{subsec_synthesis_strategy}

In this phase, we harmonize and classify the extracted data per parameter (architecture components, tools).  
To achieve this synthesis, we use \textit{card-sorting}, a lightweight, collaborative, qualitative analysis technique~\cite{Zimmermann2016}.

In particular, we use \textit{hybrid} card-sorting~\cite{hudson2014_cardsorting}, a combination of open card-sorting (where categories emerge from the data) and closed card-sorting (categories are defined beforehand based on existing taxonomies). 
We define the initial set of categories based on our background knowledge of MLOps and software architecture and iteratively refine and enrich this set. 
We conduct the card-sorting in three phases: in the preparation phase, we print all the extracted components on cards; in the execution phase, we sort the cards into meaningful categories and groups and name them; in the analysis phase, we identify the relationships between the identified components and categories. 
We also disambiguate and group together the extracted tools and derive their mapping to the synthesized architecture components. The final output is a general architecture of the MLOps workflows from a structural perspective (represented in a UML component diagram), along with a map between tools and components. Lastly, we synthesize the information about the mentioned domains for which the MLOps models are proposed.

\section{Results}\label{sec_results}

The results of our study are based on extracted and synthesized data from 43 papers published between 2020 and 2024. 
This timeframe aligns with the emergence and maturation of the MLOps discipline, which is placed in early 2019~\cite{ref_introducingMLOps_oreilly_2021}. 
We observe a peak in the publications on this topic in 2022 (19/43 papers). 

Generally, we observe that most papers (20/43) propose domain-agnostic architectures. 
At the same time, six papers focus on the domain of edge computing, three on manufacturing, and each of the remaining 14 papers focus on different domains including healthcare, psychomotor learning, etc.\footnote{To observe the complete list of domains, please refer to the replication package \cite{ref_replication_package}.}
Our data also shows that 23/43 papers are authored by academic researchers, 10/43 by industrial practitioners, and 10/43 as a collaboration between the two communities. 
This is a testament to the strong interest of \textit{both} academia and industry in MLOps.

In the remainder of this section, we provide the results of the study regarding the first and second \ac{RQ}s.

\vspace{-0.6pt}
\subsection{MLOps architecture components and their dependencies (RQ1)}

In total, we synthesized 35 unique architecture components by systematically going through all the architecture figures and descriptions contained in our 43 primary studies. 
These components are domain-agnostic, i.e., they are not tied to a specific application domain. 
We also identified domain-specific components such as \textit{IoT Sensors} (IoT domain) or \textit{User Feedback Collector} (psychomotor learning domain), but excluded them for the sake of general applicability.
We group the 35 identified architecture components into 6 categories:

\begin{itemize}[nosep]
    \item \textit{Data Curation} entails components responsible for gathering and processing data for the MLOps system.
    \item \textit{Storage and Versioning} comprises components responsible for storing, versioning, and managing the data and models in the system.
    \item \textit{ML Training} includes components responsible for training and evaluating the ML models, both in the experimentation and production phases. 
    \item \textit{CI/CD} refers to the category of components responsible for continuously building and deploying ML pipelines, models, and components. 
    \item \textit{Inference} entails the components responsible for providing predictions, making subsequent decisions, and monitoring the system.
    \item \textit{Infrastructure and Supporting Services} comprises infrastructure components that provide system support, e.g., Container Manager, Orchestrator, etc.
\end{itemize} 

Table~\ref{tbl_components_aliases} displays the 35 components by category, their responsibilities, important aliases that we identified for each component name, and the number of occurrences of each.

\begin{table}[H]
\caption{MLOps architecture components, their aliases, responsibilities, and number of occurrences in the primary studies} \label{tbl_components_aliases}
\centering
\scriptsize
\begin{tabular}{|m{0.02\textwidth}|m{0.16\textwidth}|m{0.24\textwidth}|m{0.51\textwidth}|m{0.022\textwidth}|}
\specialrule{.1em}{.06em}{.06em}
\textbf{ } & \textbf{Component Name} & \textbf{Aliases} & \textbf{Responsibilities} & \# \\
\specialrule{.1em}{.06em}{.06em}
          \parbox {25mm}{\multirow{6}{*} {\rotatebox[origin=r]{90}{ Data Curation }}}
         & Data Source & external data sources & Produces and exposes data from a real-world environment, e.g., domain events, IoT sensors, human inputs, etc. & 4 \\  \cline{2-5}
         & Data Collector &  data acquisition, data loading & Collects raw data like events from various data sources. & 10 \\ \cline{2-5}
         &  Data \newline Preprocessor & data processing, 
        data cleaning,
        data validation, 
        data curation pipeline & Validates, cleans, and prepares collected data for storing as ML training datasets. & 9  \\  \cline{1-5}

       \parbox{25mm}{\multirow{18}{*} {\centering \rotatebox[origin=r]{90}{ \centering Storage and Versioning  }}}
       
          & Dataset \newline Catalogue & -- & Stores metadata of datasets in an organized inventory, allows users to browse datasets. & 1 \\ \cline{2-5}
          & Dataset \newline Repository & data store,
        data repository, 
        data versioning & Stores and versions the datasets used for ML workflows. & 10 \\ \cline{2-5}
          & Raw Data Store & data store & Stores the raw data that are collected from sources. & 2 \\ \cline{2-5}
          & Feature Store & -- & Computes and stores reusable features, serves the computed features with low latency. & 8 \\ \cline{2-5}
          & Code \newline Repository &  source code management,
        source code repository & Stores and versions the training, deployment, and application source code. & 9 \\ \cline{2-5}
          & Model  \newline Repository & model registry,
        model store & Stores and versions the trained ML models along with basic metadata, e.g., their versions, etc. & 21 \\ \cline{2-5}
          & Artifact  \newline Repository & image repository,
        container registry & Stores a packaged or containerized ML component that incorporates an ML model for inference. & 3 \\ \cline{2-5}
          & ML Metadata Repository & experiment tracking DB,
        ML metadata store & Stores metadata related to model training for experiment tracking purposes, e.g., model performance metrics, etc. & 15 \\ \cline{2-5}
          & Feedback Database & feedback store & Stores stakeholder feedback and experiences, e.g., from domain experts or engineers, which are manually considered during iterative model development. & 2 \\ \cline{1-5}

         \parbox{25mm}{\multirow{12}{*} {\rotatebox[origin=c]{90}{ ML Training }}} 
        
          & Data Labelling Component & data annotation, ground truth annotation & Adds the ground truth labels for supervised learning models to dataset instances. & 3 \\ \cline{2-5}
          & Feature \newline Engineering Pipeline & feature selection & Selects and transforms the features of the used dataset for model training. & 4 \\ \cline{2-5}
          & ML Experiment Pipeline \newline (Offline) & ML pipeline (offline), 
        manual ML pipeline, 
        data science experiments & Develops and trains ML models at design time (more experimental and manual). & 14 \\ \cline{2-5}
        
          & ML Training Pipeline \newline (Online) &  ML pipeline (online), continuous training pipeline, incremental online learning, MLOps pipeline & Continuously trains ML models at runtime in a production environment (completely automated). & 12 \\ \cline{2-5}
          & Model  \newline Evaluator & -- & Evaluates the prediction performance of the models during training. & 4 \\ \cline{1-5}

          \parbox{25mm}{\multirow{12}{*} {\rotatebox[origin=r]{90}{ CI/CD }}}
         
           & ML Pipeline Builder & build and test pipeline,
        CI tool & Builds, tests, and packages, e.g., in containers, the code of the ML pipeline. & 2 \\ \cline{2-5}
           & ML Pipeline Deployer & pipeline deployment, 
        ML training pipeline deployment,
        CD tool & Deploys the built and packaged code of the ML pipeline to staging or production environments.  & 3 \\ \cline{2-5}
           & ML Model  \newline Deployer & model deployment,  \newline deployment pipeline & Deploys the trained model packaged with the dependencies, e.g., required libraries, preprocessing code, etc. to the production environment. & 4 \\ \cline{2-5}
           & ML Component Builder & build automation pipeline,
        continuous integration (CI) & Builds and tests ML components, i.e., deployment-ready containerized ML models wrapped in an API. & 3 \\ \cline{2-5}
           & ML Component Deployer & ML deployment, 
        continuous delivery, 
        continuous deployment, 
        CI/CD pipeline & Deploys the ML components to staging or production environments.  & 7 \\ \cline{1-5}
        
    \end{tabular}
\end{table}

\addtocounter{table}{-1}

\begin{table}[H]
\caption{MLOps architecture components, their aliases, responsibilities, and number of occurrences in the primary studies \emph{(cont.)}} 
\centering
\scriptsize
\begin{tabular}{|m{0.02\textwidth}|m{0.16\textwidth}|m{0.24\textwidth}|m{0.51\textwidth}|m{0.022\textwidth}|}
\specialrule{.1em}{.06em}{.06em}
\textbf{ } & \textbf{Component Name} & \textbf{Aliases} & \textbf{Responsibilities} & \# \\
\specialrule{.1em}{.06em}{.06em}

         \parbox{25mm}{\multirow{15}{*} {\rotatebox[origin=c]{90}{ Inference }}} 
           & Inference  \newline Service &  model inference, 
        production ML service, 
        model server & Serves the trained models to provide predictions on new data (ML component). & 10 \\ \cline{2-5}
        
           & Inference  \newline Engine &  pool inference,
        local inference engine, model serving component  & Includes an ML runtime into which trained models can be continuously deployed to serve predictions. & 9 \\ \cline{2-5}
           & Runtime \newline Model \newline Monitor &  performance monitor, 
        monitoring component,
        model runtime monitor & Continuously observes the model-serving performance and infrastructure in real-time. & 12 \\ \cline{2-5}
           & Trigger & retraining triggering webhook, 
        retraining trigger & Triggers retraining of the ML models based on predefined events and intervals or a predefined performance threshold observed via Runtime Model Monitor. & 3 \\ \cline{2-5}
           & Model \newline Comparison Runner &  model comparison runner, 
        model metrics evaluator & Compares the newly trained model to the old model and deploys the better performing one. & 2 \\ \cline{2-5}
           & Decision  \newline Processor & decision processing & Derives decisions based on the predictions of the model. The decisions are then acted upon by the actors inside or outside the system. & 2 \\ \cline{1-5}

            \parbox{25mm}{\multirow{16}{*} {\rotatebox[origin=c]{90}{  Infrastructure and Supporting Services }}}

           & Resource  \newline Manager &  resource leasing,  \newline
        model engine  & Provides foundational hardware and software computational resources. The provided computational resources can be distributed or non-distributed and scalable or non-scalable. & 5 \\ \cline{2-5}
        
           & Communication Middleware & 
         message queue,
        event streaming bus & Distributes the received requests and model predictions to resources. & 4 \\ \cline{2-5}
           & Container  \newline Manager & container service & Manages, e.g., starts and stops, the containerized ML components that are built with the ML Component Builder. & 2 \\ \cline{2-5}
           & Orchestrator & adaptive scheduler, 
        workflow orchestration,
        job tracking module  & Provides system-wide orchestration, and decides the execution schedule of multiple models balancing throughput and latency. & 5 \\ \cline{2-5}
           & Log Master &  logging, 
        info collector,
        object store, predictions store & Records and saves information regarding all the actions in the system, e.g., running the services, training, user requests, predictions, etc. & 3 \\ \cline{2-5}
           & API &  API gateway 
         & Provides interaction between the components within the platform and also between the platform and external entities. & 5 \\ \cline{2-5}
           & MLOps User  \newline Interaction Manager &  Ops dashboard,
        front-end & Provides interaction between the MLOps team and the MLOps platform. & 2 \\ \cline{1-5}
     
    \end{tabular}
\end{table}

\begin{figure}[!htb]
    \centering
    \includegraphics[width=0.92\textwidth]{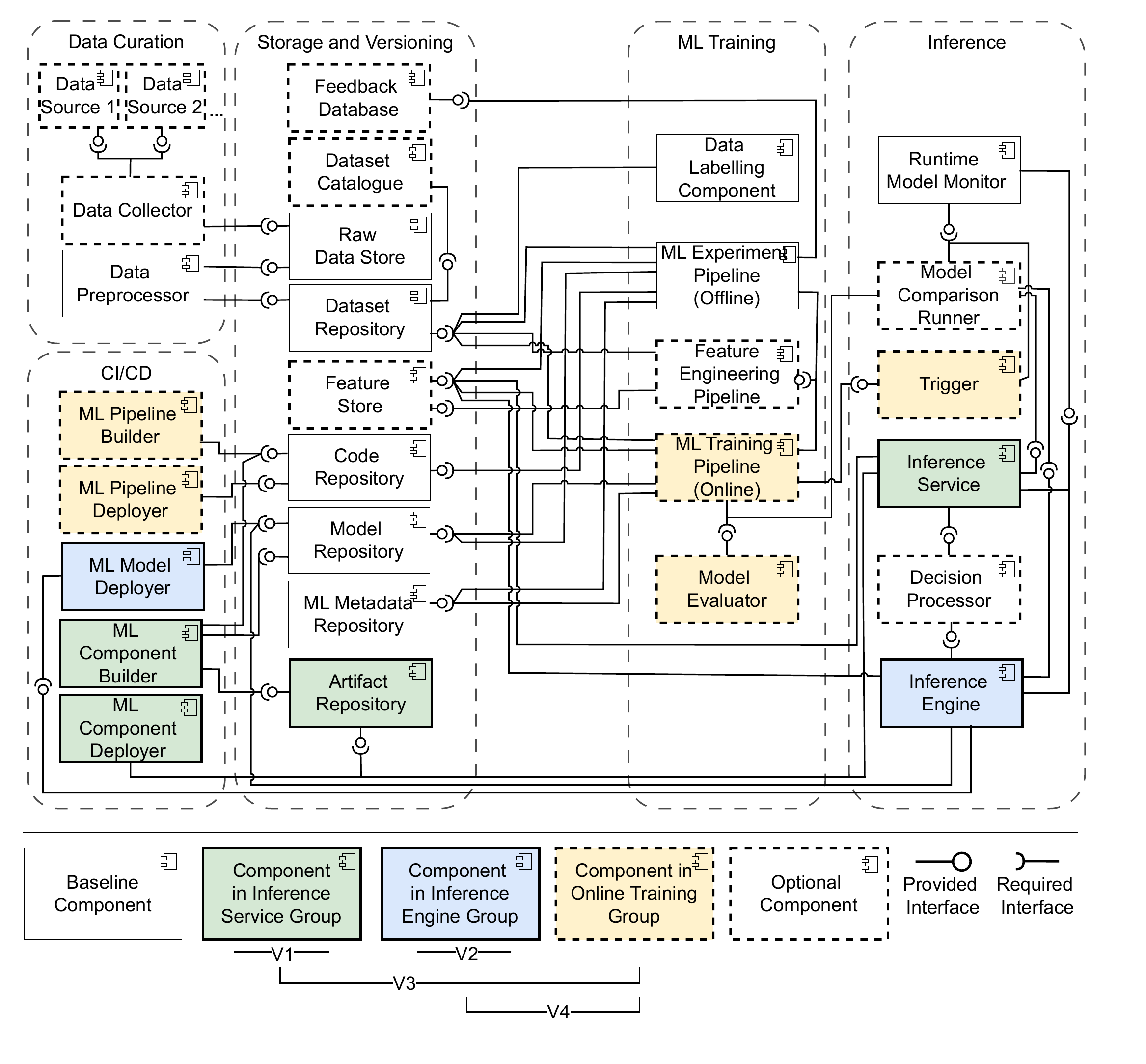}
    \caption{UML component diagram of MLOps architecture variants}
    \label{fig_component_diagram}
\end{figure}

We combined the synthesized components and their dependencies into a holistic UML component diagram depicted in~Fig.~\ref{fig_component_diagram}. 
In this diagram, we introduce five types of architecture components:
\textbf{baseline components} are mandatory components. 
These components form the baseline of the architecture (available in all variants of MLOps architectures) and need to be complemented with the components of either the \textbf{inference service} or the \textbf{inference engine} group to form a complete MLOps architecture. 
\textbf{Optional components} represent non-essential components that can be situationally useful, e.g., the \textit{Model Comparison Runner}. 
As a special type of optional components, the ones of the \textbf{online training} group can be added to an MLOps architecture, but always together as a group. 
Note that the \textit{Infrastructure and Supporting Services} components are not included in the diagram, since they provide support to the whole system. 

The baseline components include the \textit{Data Preprocessor} which reads its input from the \textit{Raw Data Store} and stores its results to the \textit{Dataset Repository}. 
The latter provides data to both the \textit{Data Labelling Component} and the \textit{ML Experiment Pipeline}. 
At the same time, the \textit{ML Experiment Pipeline} uses the datasets from the \textit{Dataset Repository} and the ML algorithms from the \textit{Code Repository} to train ML models, and then stores the trained ML models to the \textit{Model Repository}. 
After deployment, the \textit{Runtime Model Monitor} provides real-time model performance data. 

Overall, the blueprint for assembling a complete MLOps system involves the above-mentioned baseline components, as well as incorporating either the inference service or the inference engine group, and potentially the online training group and/or other optional components. 
In the following, we describe four characteristic architecture variants (V1 to V4) depicted in Fig.~\ref{fig_component_diagram}.

\textbf{V1: }
    This architecture variant describes an architecture containing an \textit{Inference Service} to serve ML models in a production environment. 
    The \textit{Inference Service} represents an \textit{ML component}~\cite{Martinez-Fernandez2022}, i.e., a containerized, deployment-ready ML model wrapped into an API that is usable for predictions. Thus, in this variant, these ML components are built, tested, and packaged by the \textit{ML Component Builder}, deployed through the \textit{ML Component Deployer}, and stored in the \textit{Artifact Repository}.

    \textbf{V2: } 
    Contrary to the \textit{Inference Service} of V1, this architecture variant involves an \textit{Inference Engine} to serve the trained ML models. 
    The \textit{Inference Engine} contains a runtime for ML models that allows the deployment of new models through the \textit{ML Model Deployer}. Hence, this variant efficiently updates only the ML model instead of always replacing the complete ML component. 
    Alternatively, the \textit{Inference Engine} can also check the \textit{Model Repository} periodically or in an event-based fashion to fetch a new model version if certain criteria are met.

    \textbf{V3: } The third variant combines the \textit{components in the online training group} with V1. The presence of \textit{Trigger} entails the presence of \textit{ML Training Pipeline (Online)}, \textit{ML Pipeline Builder}, \textit{ML Pipeline Deployer}, and the optional presence of \textit{Model Evaluator}. 
    The \textit{Trigger}, using data provided by the \textit{Runtime Model Monitor}, submits a periodic or event-based retraining request to the \textit{ML Training Pipeline}. This automatically retrains and deploys a new model in production. 

    \textbf{V4: } The fourth variant, similar to the third, combines the \textit{components in the online training group} with V2, which results in the automatic retraining and deploying of ML models in the \textit{Inference Engine}. 
    
    \textbf{Vx: } 
    In addition to these four described variants, the selection of any combination of the optional components can result in several additional variants. Table~\ref{tbl_components_aliases} can be consulted for a detailed description of all the other components. As an example, we describe the addition of the \textit{Model Comparison Runner} here. Adding this component to either V3 or V4 allows more informed model update decisions. It compares the performance of a newly trained model in the production environment to the currently deployed model and keeps the one that performs better. In the absence of this component, the newly trained model is always deployed, even if its performance would be inferior to the current ones. 

    Considering the eight optional components and four described major variants, the selection of any combination between them results in a large number of different architecture variants ($4 \times (2^8-1) = 1020$).
    Selecting the most suitable variant may depend on factors like specific design decisions, resource availability, scalability considerations, required update frequency, or technological expertise.

\subsection{Tools used to support or implement the components (RQ2)}

\label{subsec_tools_components}
\begin{figure}[htb!]
    \centering
    \includegraphics[width=0.90\textwidth]{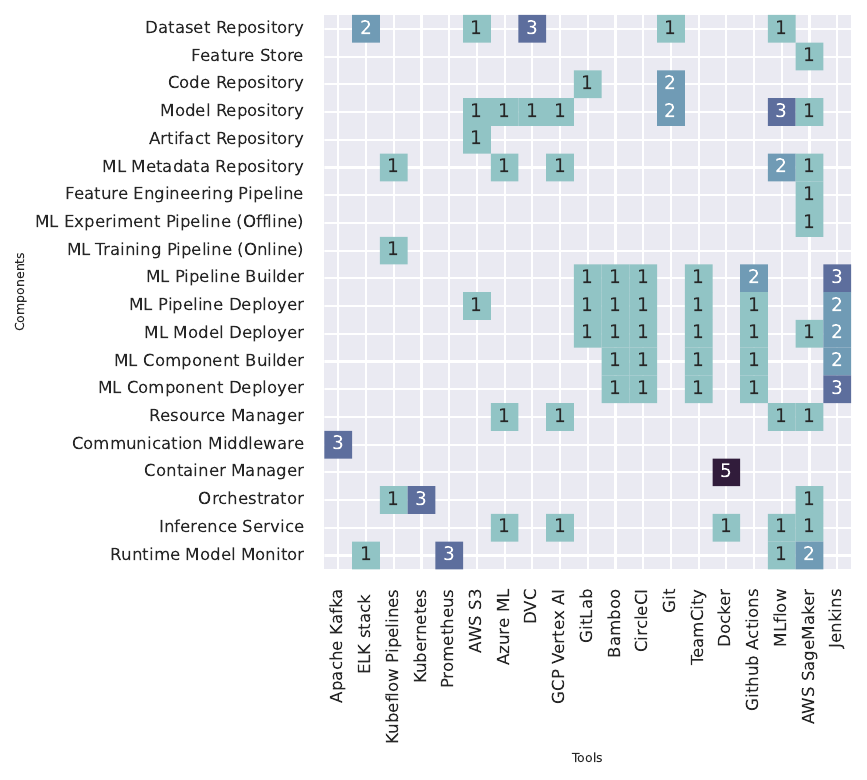}
    \caption{The most popular tools and platforms, mapped to the components}
    \label{fig_heatmap_tools}
\end{figure}

Among the 43 reviewed papers, we identified 76 tools in total.
Fig.~\ref{fig_heatmap_tools} depicts a heatmap of the tools and platforms that are mentioned at least 3 times among the papers, mapped to the components that they support. 
In the figure, it is evident that \textit{Jenkins} is the most frequently mentioned tool. 
This tool is used to support the components within the CI/CD category.
\textit{AWS SageMaker} is also one of the most popular tools. This tool is observed to support the highest number of components in the architecture of MLOps workflows (10 distinct components among 5 categories of our architecture). \textit{MLflow} follows by supporting six different components mentioned in several papers.

Within the six categories of architecture components, the greatest variety of tools is mentioned for \textit{Storage and Versioning}, followed by \textit{CI/CD}. The tools supporting the entire \textit{CI/CD} category encompass Jenkins, GitHub Actions, TeamCity, CircleCI, and Bamboo.
Regarding the components, however, the most diverse variety of tools is mentioned for \textit{Model Repository}, \textit{ML Model Deployer}, and \textit{ML Pipeline Deployer}.
There are also several components for which no tool is mentioned.
These include \textit{Dataset Catalogue}, \textit{Raw Data Store}, \textit{Feedback Database}, \textit{Data Collector}, \textit{Data Source}, \textit{Model Evaluator}, \textit{Log Master}, \textit{MLOps User Interaction Manager}, \textit{Trigger}, and \textit{Decision Service}.
The complete mapping between the tools and the components can be accessed through the replication package~\cite{ref_replication_package}. 

Lastly, some papers generally mention tools and services that support the end-to-end MLOps workflows, without mapping them to any specific component. These tools are AWS SageMaker, MLflow, Kubeflow, Weights and Biases, Clear ML, MLReef, Iguazio, Polyaxon, Vertex AI, Azure ML, and Snorkel. 

\section{Discussion}\label{sec_discussion}
In this section, we discuss the implications of the derived results of this \ac{SMS}. 
The UML component diagram and component-tool mapping provided in this study serve as valuable references for practitioners and researchers who aim to design or enhance MLOps systems. Nonetheless, the findings derived from this study suggest several noteworthy points, which we will discuss in the following.

\textbf{Complexity and mixed views in architecture diagrams.} A majority of the analyzed figures and descriptions of MLOps workflows were a combination of architecture, process, and stakeholder roles, thereby combining several different architectural views and concerns in a single diagram. Even though this approach is a common practice and makes the provided information more compact, the increased complexity reduces the clarity of the provided architectures.

\textbf{Non-standard notations to represent architectures.} 
Almost all of the analyzed figures used informal box-and-line diagrams as a notation. Among the 43 reviewed papers, only one paper (S23) uses a standard notation, the Fundamental Modelling Concepts (FMC) \cite{ref_FMC}.
Combined with the complexity of the mixed views and following the results of Warnett and Zdun~\cite{Warnett2024}, the understandability of these MLOps architecture representations is substantially impacted.

\textbf{Level of abstraction in architectures.} 
Among the extracted figures and descriptions of the papers, the level of provided details and abstraction varied over a large spectrum. During the study, we extracted and synthesized data from figures only representing as few as six high-level architecture components (S10, S43) to complex \enquote{combined figures} with over 40 concrete entities including components, actions, and stakeholder roles (S7). 
The different levels of abstraction and multitude of dependencies between components also allow many fine-grained possibilities for variations in component dependencies and interface directions.
For simplicity, we modeled only the most common dependencies in the diagram.

\textbf{Tools in place of architectural entities.}
In addition to the mixed architectural views and different levels of abstractions,  the extracted figures in some papers also included a mixture of architectural entities and the employed tools as a stand-alone entity (see, e.g., S31). This complexity in the views makes the architectural understanding and comparison difficult.

\textbf{Inconsistently named components and activities.} The terminology for the same individual components or activities could vary substantially among the papers. We identified two types of inconsistent naming: (a) some papers use common component names that imply different responsibilities, e.g., using \enquote{Feature Store} for a component that provides the training data for the ML training pipeline instead of \enquote{Data(set) Store} or using \enquote{Artifact Store} instead of \enquote{ML Model Store}, and (b) other papers use unique names to represent common ML-based software components, e.g., \enquote{knowledge base manager} for a component that 
stores and versions the trained ML models (\textit{Model Repository} in our synthesis).
These inconsistencies highlight why communication and collaboration in MLOps projects is often difficult.

\textbf{Domain-specific architecture components.}
In almost 50\% of the papers, the provided architectures are domain-specific and therefore contain components that are only situationally applicable or focused on a specific application scenario. For example, the presence of \enquote{additive manufacturing} components in S36, or \enquote{Blockchain}-related components in S40. In the analysis, we generalized these components to a relevant category or discarded the components that represented a very specialized entity.

\textbf{End-to-end tools only mentioned for a certain set of components.} 
A tool-related observation is that most of the end-to-end MLOps tools like MLflow, Kubeflow, or AWS SageMaker are recommended only for a subset of components, rather than for the entire workflow. For example, in the reviewed papers, MLflow is mentioned only for components in the \textit{Storage and Versioning},  \textit{Inference}, and \textit{Infrastructure and Supporting Services} among our categories. Interestingly, this tool is not mentioned for any component in the \textit{ML Training} category.

\section{Threats to validity}
Following the categorization by Wohlin et al.~\cite{wohlin2012experimentation} and the checklist by Ampatzoglou~\cite{ampatzoglou2019threatsinSE}, we outline the threats that may have affected the validity of our research and outline the actions that we take to mitigate the threats.

Threats to \textit{internal validity} undermine the conclusion about a possible causal relationship between the study and the outcome~\cite{wohlin2012experimentation}. 
To mitigate this threat, we assign two researchers for the study selection and data extraction phases, who perform the selection and extraction independently, and then three researchers discuss the results in consensus meetings. 
Another possible threat to internal validity in this study stems from the selection of papers from various domains. Varying author expertise across fields, particularly outside ML-based systems, may influence the accuracy of the architecture and process descriptions in papers, which influences our data extraction. To mitigate this threat, we have mapped every component and activity name and description to the authors' intentions, e.g., in some papers authors refer to a dataset repository as a feature store. 

\textit{External validity} is concerned with the generalization extent of the findings of the study~\cite{wohlin2012experimentation}. 
A possible threat to the external validity of our study is the selection of the papers among the peer-reviewed resources, which may limit our initial set of papers. This is a research design decision to ensure the reliability of the selected papers. 
We mitigate this threat by applying our search query on Google Scholar as a meta-engine, which results in papers from different venues, as well as applying one round of bidirectional snowballing. 
However, since we focus on peer-reviewed scientific papers, we might miss works from the many practitioners that certainly focus on this very popular topic. 
This may introduce a threat to the generalizability of the results.

\textit{Reliability} is concerned with the extent to which the data and the analysis are dependent on the specific researchers~\cite{ampatzoglou2019threatsinSE,runeson2009guidelinesinSE}. 
A possible threat in our study concerning reliability can be the authors' bias in synthesizing data due to the nature of the extracted data and the different levels of abstraction of the architecture in the papers.
To mitigate this threat, three authors participate in analyzing and synthesizing the data.

\section{Conclusions}
\label{sec_conclusion}
Based on an \ac{SMS} with 43 scientific papers, we synthesized architectural MLOps components, their dependencies, and tools to implement them.
Furthermore, we combined these components into a holistic MLOps architecture and discussed several architectural variants that emerge from the literature.
Our results contribute to understanding the architecture aspects of MLOps systems and potentially support communication in this complex and still maturing domain.

Regarding future work, a structural perspective is not the only architectural view that is important in MLOps.
Therefore, we plan to synthesize a process view of MLOps with a similar research design and to map activities to MLOps roles.
Moreover, synthesizing architectural decisions, best practices, and antipatterns in this domain may also support practitioners, which we plan to provide via our long-term goal, a reference architecture for MLOps.
In the end, practitioners could be best supported if we could clearly link architectural MLOps variants to functional and quality requirements, so that practitioners can easily choose the variant that best suits their needs.
To allow such endeavors and to increase transparency, we share our research artifacts on Zenodo~\cite{ref_replication_package}.

\section*{Data Availability}
The data and artifacts of this study are available as a replication package~\cite{ref_replication_package}.

\begin{credits}
\subsubsection{\ackname} 
This research is supported by ExtremeXP, a project co-funded by the European Union Horizon Programme under Grant Agreement No. 101093164.
\end{credits}

\bibliographystyle{splncs04}
\bibliography{bibliography}

\begin{thebibliography}{10}
\providecommand{\url}[1]{\texttt{#1}}
\providecommand{\urlprefix}{URL }
\providecommand{\doi}[1]{https://doi.org/#1}

\bibitem{ref_replication_package}
Amou~Najafabadi, F., Bogner, J., Gerostathopoulos, I., Lago, P.: {An Analysis of MLOps Architectures: A Systematic Mapping Study}. [Data set], Zenodo (2024), \url{https://doi.org//10.5281/zenodo.11067770}

\bibitem{ampatzoglou2019threatsinSE}
Ampatzoglou, A., Bibi, S., Avgeriou, P., Verbeek, M., Chatzigeorgiou, A.: Identifying, categorizing and mitigating threats to validity in software engineering secondary studies. Information and Software Technology  \textbf{106},  201--230 (2019)

\bibitem{Bass2015}
Bass, L., Weber, I., Zhu, L.: {DevOps}: {A} {Software} {Architect}'s {Perspective}. Addison-Wesley Professional, 1st edn. (2015)

\bibitem{ref_FMC}
The fundamental modeling concepts, \url{http://www.fmc-modeling.org}, online.

\bibitem{hudson2014_cardsorting}
Hudson, W.: Card sorting. The Encyclopedia of Human-Computer Interaction  (2014), \url{https://www.interaction-design.org/literature/book/the-encyclopedia-of-human-computer-interaction-2nd-ed/card-sorting}, accessed Mar. 31, 2024

\bibitem{idowu_asset_2021}
Idowu, S., Strüber, D., Berger, T.: Asset management in machine learning: a survey. In: Proceedings of the 43rd {International} {Conference} on {Software} {Engineering}: {Software} {Engineering} in {Practice}. pp. 51--60. {ICSE}-{SEIP} '21, IEEE Press (2021)

\bibitem{John2021}
John, M.M., Olsson, H.H., Bosch, J.: Towards {MLOps}: {A} {Framework} and {Maturity} {Model}. In: 2021 47th {Euromicro} {Conference} on {Software} {Engineering} and {Advanced} {Applications} ({SEAA}). pp.~1--8. IEEE (2021)

\bibitem{keele2007_SLRguidelines}
Keele, S., Kitchenham, B., et~al.: Guidelines for performing systematic literature reviews in software engineering (2007)

\bibitem{kolar2023barriers_MLOps}
Kolar~Narayanappa, A., Amrit, C.: An analysis of the barriers preventing the implementation of mlops. In: International Working Conference on Transfer and Diffusion of IT. pp. 101--114. Springer (2023)

\bibitem{kolltveit2022operationalizingMLModels}
Kolltveit, A.B., Li, J.: Operationalizing machine learning models: A systematic literature review. In: Proceedings of the 1st Workshop on Software Engineering for Responsible AI. pp.~1--8 (2022)

\bibitem{kreuzberger_machine_2023}
Kreuzberger, D., Kühl, N., Hirschl, S.: Machine {Learning} {Operations} ({MLOps}): {Overview}, {Definition}, and {Architecture}. IEEE Access  \textbf{11},  31866--31879 (2023)

\bibitem{kumara2023requirements_ra_mlops}
Kumara, I., Arts, R., Di~Nucci, D., Van Den~Heuvel, W.J., Tamburri, D.A.: Requirements and reference architecture for mlops: Insights from industry. Authorea Preprints  (2023)

\bibitem{lima2022mlops_SLR}
Lima, A., Monteiro, L., Furtado, A.P.: Mlops: Practices, maturity models, roles, tools, and challenges-a systematic literature review. ICEIS (1) pp. 308--320 (2022)

\bibitem{ref_introducingMLOps_oreilly_2021}
Mark~Treveil, t.D.T.: Introducing MLOps. O'Reilly Media, Inc (2021)

\bibitem{Martinez-Fernandez2022}
{Mart{\'i}nez-Fern{\'a}ndez}, S., Bogner, J., Franch, X., Oriol, M., Siebert, J., Trendowicz, A., Vollmer, A.M., Wagner, S.: Software {{Engineering}} for {{AI-Based Systems}}: {{A Survey}}. ACM Transactions on Software Engineering and Methodology  \textbf{31}(2) (2022)

\bibitem{mboweni2022SLR_MLDevOps}
Mboweni, T., Masombuka, T., Dongmo, C.: A systematic review of machine learning devops. In: 2022 international conference on electrical, computer and energy technologies (ICECET). pp.~1--6. IEEE (2022)

\bibitem{nakagawa2023reference-architecture}
Nakagawa, E.Y., Antonino, P.O.: Reference Architectures for Critical Domains: Industrial Uses and Impacts. Springer Nature (2023)

\bibitem{petersen_2015_SMSguidelines}
Petersen, K., Vakkalanka, S., Kuzniarz, L.: Guidelines for conducting systematic mapping studies in software engineering: An update. Information and Software Technology  \textbf{64},  1--18 (2015)

\bibitem{recupito2022MLOpsTools_MLR}
Recupito, G., Pecorelli, F., Catolino, G., Moreschini, S., Di~Nucci, D., Palomba, F., Tamburri, D.A.: A multivocal literature review of mlops tools and features. In: 2022 48th Euromicro Conference on Software Engineering and Advanced Applications (SEAA). pp. 84--91. IEEE (2022)

\bibitem{runeson2009guidelinesinSE}
Runeson, P., H{\"o}st, M.: Guidelines for conducting and reporting case study research in software engineering. Empirical software engineering  \textbf{14},  131--164 (2009)

\bibitem{testi2022SLR_mlops_taxonomy}
Testi, M., Ballabio, M., Frontoni, E., Iannello, G., Moccia, S., Soda, P., Vessio, G.: Mlops: A taxonomy and a methodology. IEEE Access  \textbf{10},  63606--63618 (2022)

\bibitem{warnett_architectural_2022}
Warnett, S.J., Zdun, U.: Architectural {Design} {Decisions} for {Machine} {Learning} {Deployment}. In: 2022 {IEEE} 19th {International} {Conference} on {Software} {Architecture} ({ICSA}). pp. 90--100 (2022)

\bibitem{Warnett2022}
Warnett, S.J., Zdun, U.: Architectural {Design} {Decisions} for the {Machine} {Learning} {Workflow}. Computer  \textbf{55}(3),  40--51 (2022), publisher: IEEE

\bibitem{Warnett2024}
Warnett, S.J., Zdun, U.: On the {Understandability} of {MLOps} {System} {Architectures}. IEEE Transactions on Software Engineering pp. 1--25 (2024)

\bibitem{Wohlin2014_snowballing}
Wohlin, C.: Guidelines for snowballing in systematic literature studies and a replication in software engineering. In: Proceedings of the 18th International Conference on Evaluation and Assessment in Software Engineering. EASE '14, ACM (2014)

\bibitem{wohlin2012experimentation}
Wohlin, C., Runeson, P., H{\"o}st, M., Ohlsson, M.C., Regnell, B., Wessl{\'e}n, A.: Experimentation in software engineering. Springer Science \& Business Media (2012)

\bibitem{Zimmermann2016}
Zimmermann, T.: {Card-sorting: From text to themes}. In: Perspectives on Data Science for Software Engineering, pp. 137--141. Elsevier (2016)

\end{thebibliography}

\end{document}